# Diffusion $L_0$-norm constraint improved proportionate LMS algorithm for sparse distributed estimation

Zongsheng Zheng, Zhigang Liu

## Abstract

To exploit the sparsity of the considered system, the diffusion proportionate-type least mean square (PtLMS) algorithms assign different gains to each tap in the convergence stage while the diffusion sparsity-constrained LMS (ScLMS) algorithms pull the components towards zeros in the steady-state stage. In this paper, by minimizing a differentiable cost function that utilizes the Riemannian distance between the updated and previous weight vectors as well as the $L_0$ norm of the weighted updated weight vector, we propose a diffusion $L_0$-norm constraint improved proportionate LMS ($L_0$-IPLMS) algorithm, which combines the benefits of the diffusion PtLMS and diffusion ScLMS algorithms and performs the best performance among them. Simulations in a system identification context confirm the improvement of the proposed algorithm.

## 1. Introduction

To estimate some parameters of interest from the data collected at nodes distributed over a geographic region, the distributed estimation was introduced [1-5]. In the distributed estimation, every node in the network communicates with a subset of the nodes, and the estimation is performed at each node in the network. Several strategies have been proposed for sequential data processing over networks, including the consensus [3], incremental [1], and diffusion [4] strategies.

The diffusion strategy is particularly attractive due to its enhanced adaptation performance and wide stability range [5]. It uses the subset of neighbours to communicate, and therefore requires low computational complexity and owns stable behaviour in real-time adaptation. The diffusion least mean square (LMS) algorithm was first proposed in [2]. In [4], a general form of diffusion LMS algorithms was presented in which the adapt-then-combine (ATC) and combine-then-adapt (CTA) versions of diffusion LMS algorithms were formulated.

In many situations, the parameter of interest is sparse, which means that it has only a few relatively large components and the other components are negligible. To exploit information on sparsity, two classes of diffusion algorithms have

The authors are with the School of Electrical Engineering, Southwest Jiaotong University, Chengdu, 610031, China. E-mail addresses: bk20095185@my.swjtu.edu.cn (Z. Zheng), liuzg_cd@126.com (Z. Liu).

been presented: diffusion sparsity-constrained LMS (ScLMS) [6, 7] and diffusion proportionate-type LMS (PtLMS) [8, 9] algorithms. In the diffusion PtLMS algorithms, each coefficient of the filter is updated independently by adjusting the gain in proportion to the magnitude of the estimated filter coefficient. In this way, the larger coefficient receives larger increment, thus increasing the convergence rate of that coefficient. The diffusion ScLMS algorithms are derived by adding the sparsity constraints to the cost function. This technique is equivalent to adding a zero-attracting term in the iteration of the LMS-based algorithm, which accelerates the convergence rates of the zero or near-zero components.

To enhance the detection of sparsity in the underlying system model, we propose a diffusion $L_0$-norm constraint improved proportionate LMS ($L_0$-IPLMS) algorithm by minimizing a differentiable cost function that utilizes the Riemannian distance between the updated and previous weight vectors as well as the $L_0$ norm of the weighted updated weight vector. The diffusion $L_0$-IPLMS algorithm combines the benefits of the diffusion PtLMS and diffusion ScLMS algorithms and performs the best performance among them.

## 2. Diffusion $L_0$-IPLMS algorithm

Let us consider a connected network with $N$ nodes. At each time $i$, each node $k$ gets a noisy measurement $d_k(i)$ and input vector $\boldsymbol{u}_{k,i} = [u_{k,i}, u_{k,i-1}, ..., u_{k,i-L+1}]^T$ with length $L$, which are related via the following linear model

$$d_k(i) = \boldsymbol{u}_{k,i}\boldsymbol{w}_o + v_k(i) \tag{1}$$

where $\boldsymbol{w}_o$ is the unknown $L$-dimensional parameter vector of interest and $v_k(i)$ is the measurement noise with variance $\sigma_{v,k}^2$.

Define the a priori and intermediate a posteriori error of node $k$ by

$$e_k(i) = d_k(i) - \boldsymbol{u}_{k,i}^T \boldsymbol{w}_{k,i-1} \tag{2}$$

$$\varepsilon_k(i) = d_k(i) - \boldsymbol{u}_{k,i}^T \boldsymbol{\psi}_{k,i} \tag{3}$$

where $\boldsymbol{w}_{k,i-1}$ is the estimate of $w^o$ at node $k$ and time $i$-1, $\boldsymbol{\psi}_{k,i}$ is an intermediate estimate of $w^o$ at node $k$ and time $i$-1.

According to [10], an efficient adaptive algorithm must be conservative (avoid radical changes of estimate from one iteration to the next) and corrective (decrease the difference between the measurement and output). Toward this end, we consider the following cost function:

$$J_{k,i} = \varepsilon_k^2(i) + \left\|\boldsymbol{\psi}_{k,i} - \boldsymbol{w}_{k,i-1}\right\|_{\boldsymbol{R}_{k,i}}^2 + \gamma \boldsymbol{R}_{k,i} \left\|\boldsymbol{\psi}_{k,i}\right\|_0 \tag{4}$$

where $\gamma$ is a positive constant, $\boldsymbol{R}_{k,i}$ is a positive definite matrix whose entries depend on $\boldsymbol{w}_{k,i-1}$, i.e., $\boldsymbol{R}_{k,i}$ is an $L \times L$ Riemannian metric tensor, then the term $\left\|\boldsymbol{\psi}_{k,i} - \boldsymbol{w}_{k,i-1}\right\|^2_{\boldsymbol{R}_{k,i}} = [\boldsymbol{\psi}_{k,i} - \boldsymbol{w}_{k,i-1}]^T \boldsymbol{R}_{k,i} [\boldsymbol{\psi}_{k,i} - \boldsymbol{w}_{k,i-1}]$ represents the Riemannian distance between $\boldsymbol{\psi}_{k,i}$ and $\boldsymbol{w}_{k,i-1}$, and it ensures the conservativeness of the algorithm. The term $\varepsilon_k^2(i)$ minimizes the squared error of the measurement and output, and it is usually used in the cost function of the adaptive algorithm. Given that $\|\cdot\|_0$ is the $L_0$ norm that counts the number of nonzero entries in $\boldsymbol{\psi}_{k,i}$, $\|\boldsymbol{\psi}_{k,i}\|_0$ owns the ability of accelerating the convergence rates of the zero or near-zero components.

To exploit the sparsity of the impulse response, we usually set $\boldsymbol{R}_{k,i} = \boldsymbol{G}_{k,i}^{-1}$, where $\boldsymbol{G}_{k,i}$ is a diagonal matrix $\boldsymbol{G}_{k,i} = \mathrm{diag}\{g_{k,0}(i), g_{k,1}(i), ..., g_{k,L-1}(i)\}$ called proportionate matrix. For the diagonal element $g_{k,l}(i)$ with $0 \le l \le L-1$, there are several choices in [11, 12]. Among these, the following equation is one of the most attractive choices because of its robustness to impulse responses with different sparseness degrees [12]

$$g_{k,l}(i) = \frac{1-\alpha}{2} + (1+\alpha)\frac{L\left|w_{k,l}(i)\right|}{2\sum_{j=0}^{L-1}\left|w_{k,j}(i)\right| + \varepsilon}, \quad l = 0,1,...,L-1 \tag{5}$$

where $-1 \le \alpha \le 1$ and $\varepsilon$ is a small positive constant to avoid division by zero.

Since the $L_0$ norm minimization is a Non-Polynomial (NP) hard problem, a continuous function is usually used to approximate the $L_0$ norm [13]

$$\|\boldsymbol{\psi}_{k,i}\|_0 \approx F_\varphi(\boldsymbol{\psi}_{k,i}) = \sum_{l=0}^{L-1}\left(1 - e^{-\varphi|\psi_{k,l}(i)|}\right), \quad \varphi > 0. \tag{6}$$

Taking into consideration the discussion above, the cost function can be rewritten as

$$J_{k,i} = \varepsilon_k^2(i) + \left\|\boldsymbol{\psi}_{k,i} - \boldsymbol{w}_{k,i-1}\right\|^2_{\boldsymbol{G}_{k,i}^{-1}} + \gamma \boldsymbol{G}_{k,i}^{-1} F_\varphi(\boldsymbol{\psi}_{k,i}) \tag{7}$$

Taking the derivative of (7) with respect to the intermediate estimate $\boldsymbol{\psi}_{k,i}$, we have

$$\frac{\partial J_{k,i}}{\partial \boldsymbol{\psi}_{k,i}} = -2\boldsymbol{u}_{k,i}\varepsilon_k(i) + 2\boldsymbol{G}_{k,i}^{-1}(\boldsymbol{\psi}_{k,i} - \boldsymbol{w}_{k,i-1}) + \gamma \boldsymbol{G}_{k,i}^{-1} f_\varphi(\boldsymbol{\psi}_{k,i}) \tag{8}$$

where $f_\varphi(\boldsymbol{\psi}_{k,i}) = \partial F_\varphi(\boldsymbol{\psi}_{k,i}) / \partial \boldsymbol{\psi}_{k,i} = [f_\varphi(\psi_{k,l}(i)), f_\varphi(\psi_{k,1}(i))..., f_\varphi(\psi_{k,L-1}(i))]$ and $f_\varphi(\psi_{k,l}(i)) = \varphi \,\mathrm{sgn}(\psi_{k,l}(i)) e^{-\varphi|\psi_{k,l}(i)|}$ for $l = 0,1,...,L-1$.

Setting the derivative in (8) equal to zero, we get

$$\boldsymbol{\psi}_{k,i} = \boldsymbol{w}_{k,i-1} + \boldsymbol{G}_{k,i}\boldsymbol{u}_{k,i}\varepsilon_k(i) - \frac{1}{2}\gamma f_\varphi(\boldsymbol{\psi}_{k,i}) \tag{9}$$

Since the intermediate a posteriori error $\varepsilon_k(i)$ depends on the intermediate estimate $\boldsymbol{\psi}_{k,i}$ which is not accessible before the current update, it is reasonable to approximate it with the a priori error $e_k(i)$. We replace $f_\varphi(\boldsymbol{\psi}_{k,i})$ by

$f_{\varphi}(\boldsymbol{w}_{k,i-1})$ for the same reason. Then, (9) can be further expressed as

$$\boldsymbol{\psi}_{k,i} = \boldsymbol{w}_{k,i-1} + \mu \boldsymbol{G}_{k,i}\boldsymbol{u}_{k,i}e_k(i) - \rho f_{\varphi}(\boldsymbol{w}_{k,i-1}) \quad (10)$$

where we added the step size $\mu$ to control the convergence of the algorithm, and $\rho = \frac{1}{2}\gamma\mu$.

The diffusion algorithm performs the estimation with two steps: adaptation and combination. According to the order of these two steps, the diffusion algorithm is classified into the combine-then-adapt (CTA) and combine-then-adapt (CTA) diffusion algorithms [4]. We adopt the ATC diffusion algorithm which achieves lower steady-state error than the CTA diffusion algorithm [4]. The diffusion $L_0$-IPLMS algorithm can be described as

$$\begin{cases} \boldsymbol{\psi}_{k,i} = \boldsymbol{w}_{k,i-1} + \mu \boldsymbol{G}_{k,i}\boldsymbol{u}_{k,i}e_k(i) - \rho f_{\varphi}(\boldsymbol{w}_{k,i-1}) \\ \boldsymbol{w}_{k,i} = \sum_{n \in N_k} c_{n,k}\boldsymbol{\psi}_{n,i} \end{cases} \quad (11)$$

where $N_k$ denotes the set of nodes in the neighborhood of node $k$ including itself, $\{c_{l,k}\}$ are real, non-negative, and satisfy $\sum_{l=1}^{N} c_{l,k} = 1$.

**Remark #1:** From the diffusion $L_0$-IPLMS algorithm, some existing algorithms can be obtained below

1) If the proportionate matrix becomes identity matrix, i.e., $\boldsymbol{G}_{k,i} = \mathbf{I}_L$, the Riemannian distance $\|\boldsymbol{\psi}_{k,i} - \boldsymbol{w}_{k,i-1}\|^2_{\boldsymbol{G}^{-1}_{k,i}}$ becomes the Euclidean distance $\|\boldsymbol{\psi}_{k,i} - \boldsymbol{w}_{k,i-1}\|^2_2$, and then the diffusion $L_0$-IPLMS algorithm reduces to the diffusion $L_0$-LMS algorithm;

2) When the weight given to the $L_0$ norm penalty becomes zero, i.e., $\rho = 0$, the diffusion $L_0$-IPLMS algorithm reduces to the diffusion IPLMS algorithm;

3) The diffusion $L_0$-IPLMS algorithm becomes the diffusion LMS algorithm when $\boldsymbol{G}_{k,i} = \mathbf{I}_L$ and $\rho = 0$.

**Remark #2:** Our work can be considered as a generalization of the diffusion PtLMS and diffusion ScLMS algorithms.

1) For diffusion ScLMS algorithms (i.e., $\boldsymbol{G}_{k,i} = \mathbf{I}_L$), different kinds of ScLMS algorithms can be obtained by replacing the $L_0$ norm with the $L_1$ norm, $L_p$ norm (0 < p < 1) and so on.

2) For diffusion PtLMS algorithms (i.e., $\rho = 0$), by choosing different proportionate matrix $\boldsymbol{G}_{k,i}$ (that is different diagonal element $g_{k,l}(i)$), lots of PtLMS algorithms can be achieved, such as proportionate LMS, sparseness-controlled proportionate LMS, μ-law proportionate LMS and individual-activation-factor proportionate LMS.

## 3. Simulation results

In this section, we present some numerical examples to illustrate the performance of the diffusion $L_0$-IPLMS algorithm.

We consider a connected network composed of 20 nodes, as illustrated in Fig. 1. The uniform rule is used for $\{c_{l,k}\}$. The regressors are zero-mean white Gaussian distributed with covariance matrices $R_{u,k} = \sigma_{u,k}^2 I_M$, with $\sigma_{u,k}^2$ shown in Fig. 2. The background noise power and corresponding SNR of each node are also depicted in Fig. 2. The network mean-square deviation (MSD), which is defined as $10\log_{10}[\frac{1}{N}\sum_{l=1}^{N}\|w_l - w^o\|_2^2]$, is used to measure the performance of the algorithms. The simulation results are obtained by ensemble averaging over 100 trials. In the following simulations, we set $\varepsilon = 0.01$, $\alpha = 0$ [12], and $\varphi = 5$ [13].

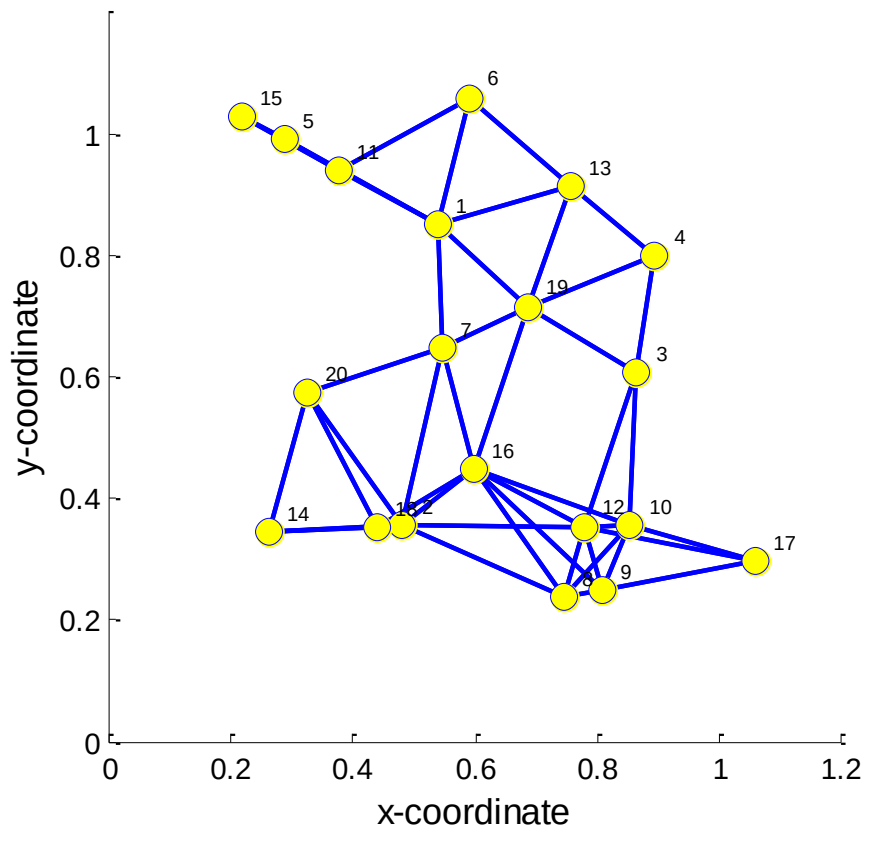

Fig. 1. Network topology.

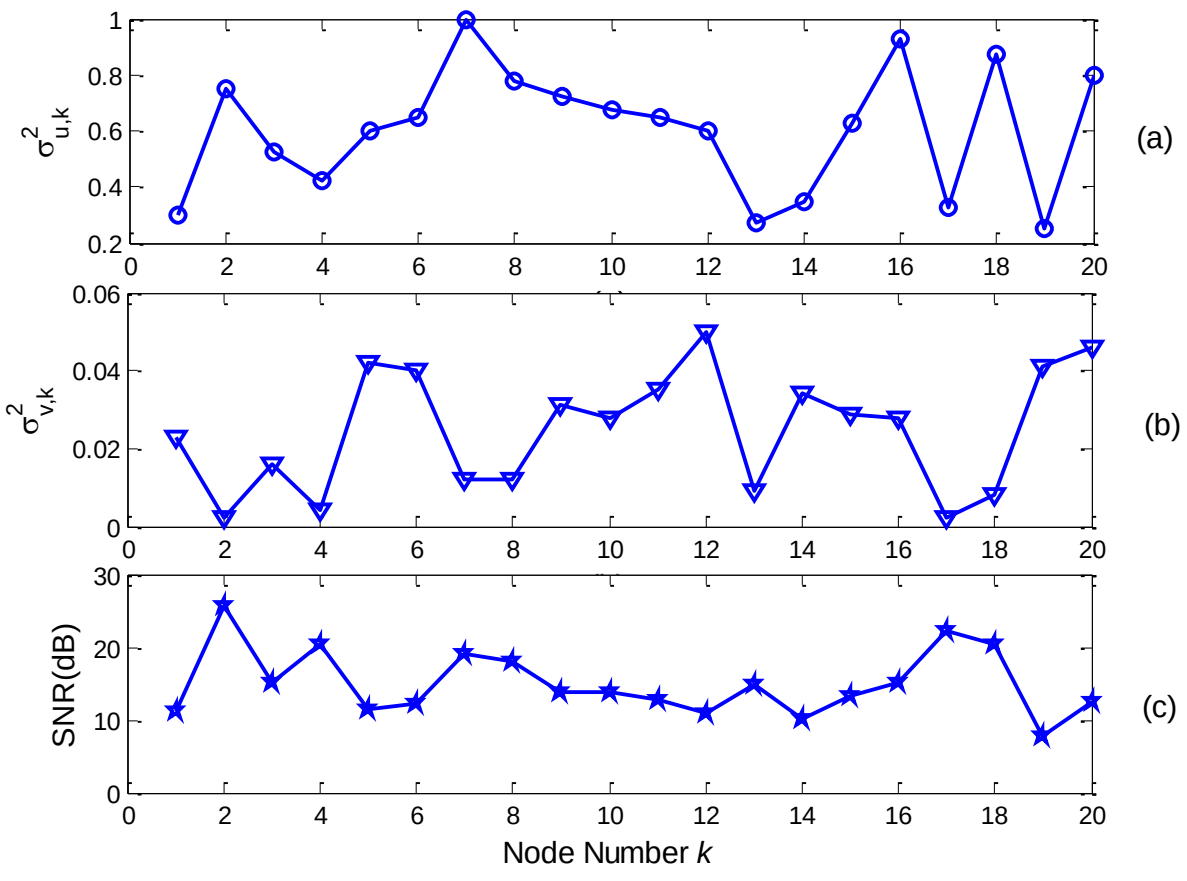

Fig. 2. (a) Regressor variances $\sigma_{u,k}^2$, (b) noise variances $\sigma_{v,k}^2$, (c) signal-to-noise ratio.

First, we compare the performance of the diffusion $L_0$-IPLMS algorithm with that of the diffusion LMS, diffusion $L_0$-LMS, and diffusion IPLMS algorithms for sparse distributed estimation in Fig. 3, where the unknown vector of interest $w^o = [1, 0,.., 0,1]^T / \sqrt{2}$ ($M = 50$) is considered. For a fair comparison, the step size $\mu$ is determined in such a way that all algorithms have the same initial convergence rate, and the parameter $\rho$ is set according to [13]. Compared with the diffusion LMS algorithm, the diffusion $L_0$-LMS and diffusion IPLMS algorithms obtain better performance in terms of steady-state misalignment. The proposed diffusion $L_0$-IPLMS algorithm shows the best performance among them.

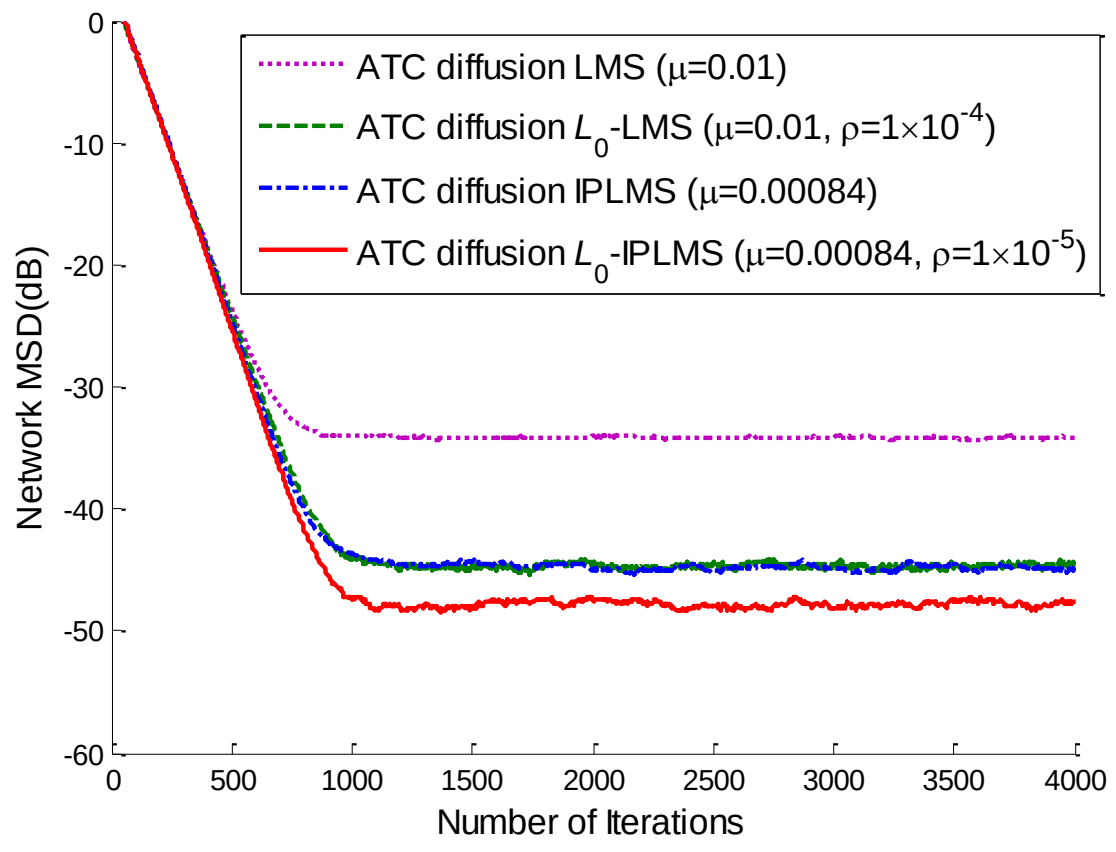

Fig. 3. Network MSD comparison among diffusion LMS, diffusion IPLMS and diffusion $L_0$-IPLMS algorithms.

In Fig. 4, we test the performance of the diffusion $L_0$-IPLMS algorithm for systems with different sparsity ratios. As can be seen, the diffusion $L_0$-IPLMS algorithm outperforms the diffusion $L_0$-LMS and diffusion IPLMS algorithms for both sparse and non-sparse systems.

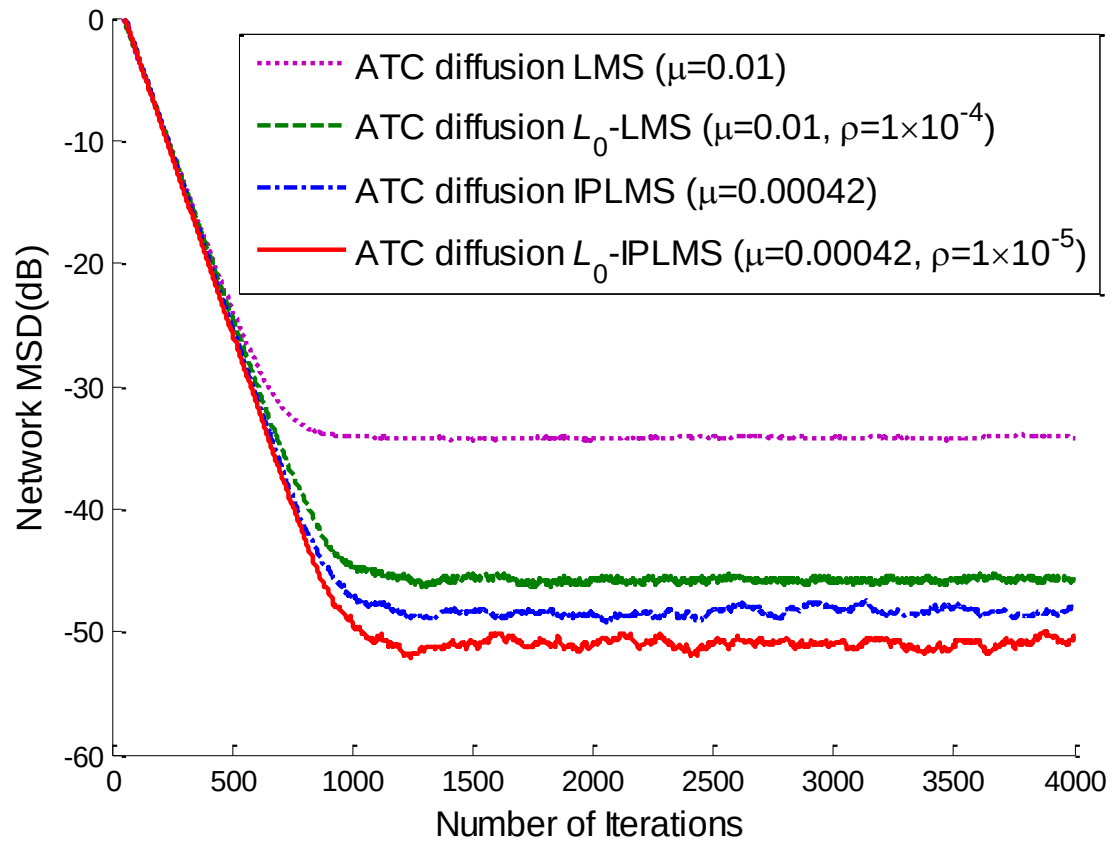

(a)

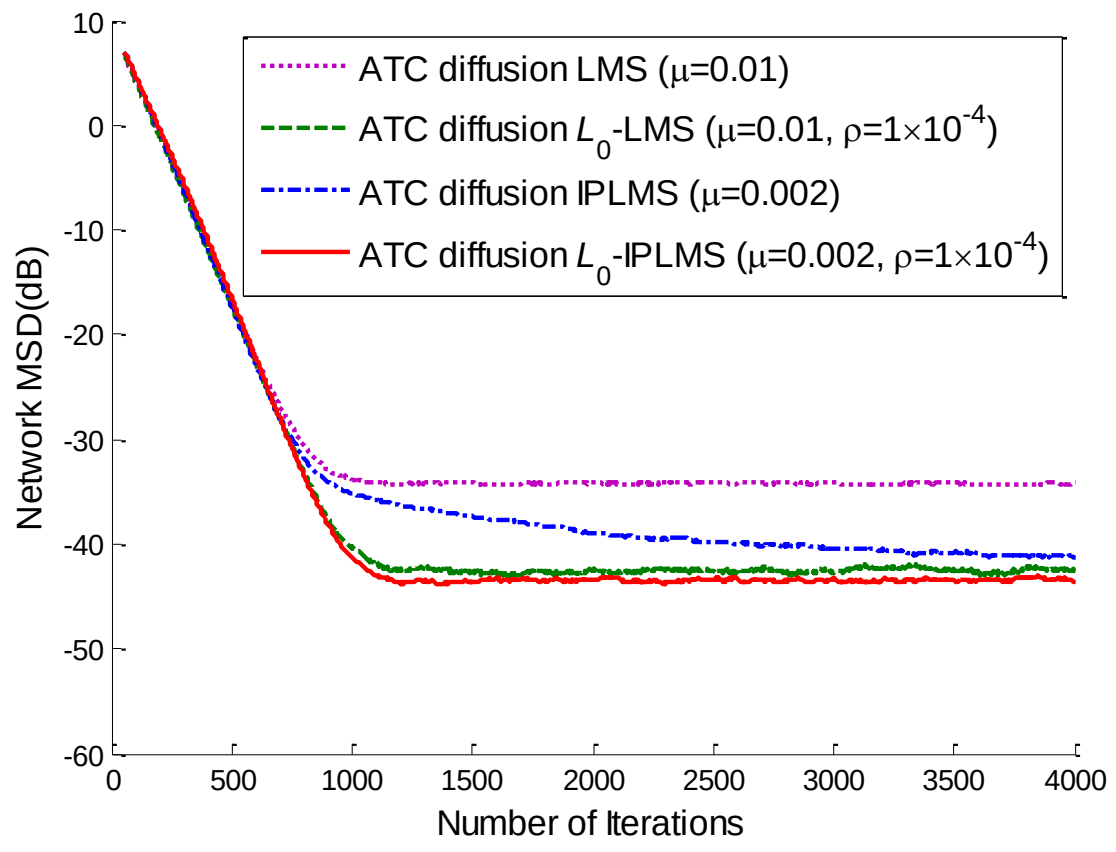

(b)

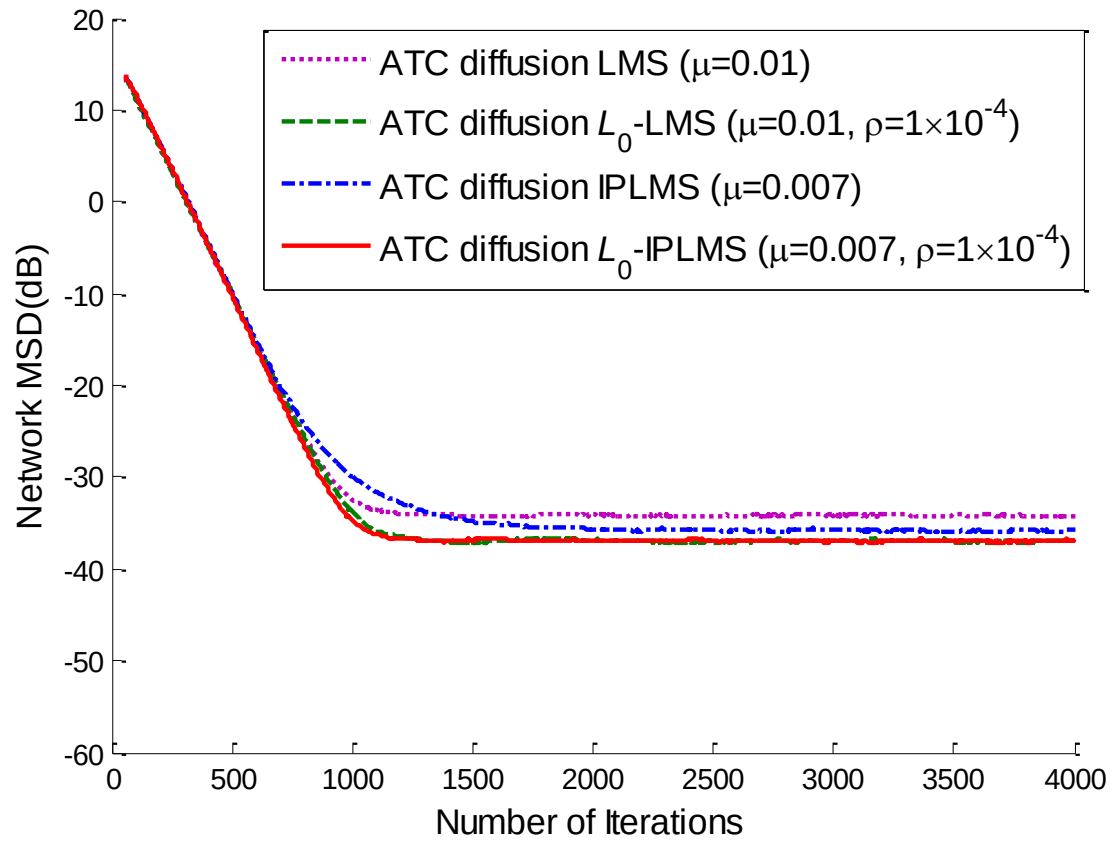

(c)

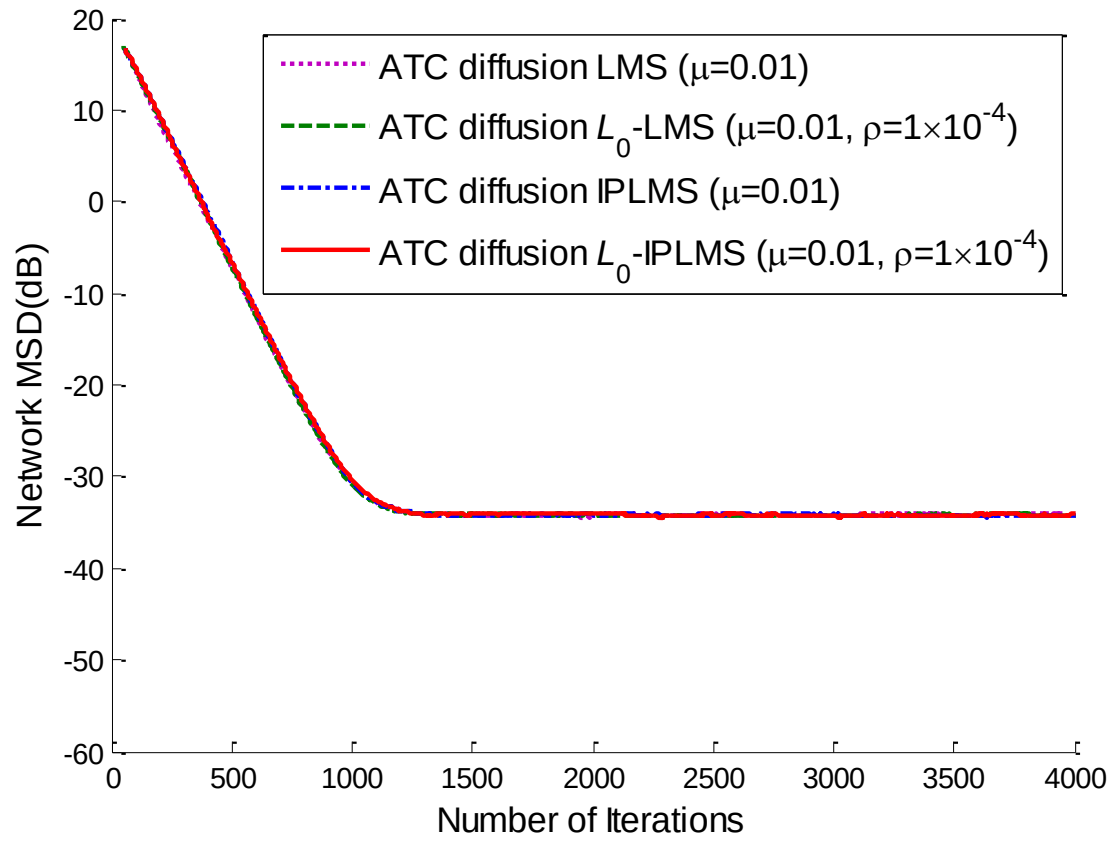

(d)

Fig. 4. Network MSD comparison among diffusion LMS, diffusion IPLMS and diffusion $L_0$-IPLMS algorithms. (a) sparsity ratio = 1/50, (b) sparsity ratio = 5/50, (c) sparsity ratio = 25/50, (c) sparsity ratio = 50/50.

## 4. Conclusions

To exploit the sparsity of the considered system, the diffusion proportionate-type least mean square (PtLMS) algorithms assign different gains to each tap in the convergence stage while the diffusion sparsity-constrained LMS (ScLMS) algorithms pull the components towards zeros in the steady-state stage. In this paper, by minimizing a differentiable cost function that utilizes the Riemannian distance between the updated and previous weight vectors as well as the $L_0$ norm of the weighted updated weight vector, we propose a diffusion $L_0$-norm constraint improved proportionate LMS ($L_0$-IPLMS) algorithm, which combines the benefits of the diffusion PtLMS and diffusion ScLMS algorithms and performs the best performance among them. Simulations in a system identification context confirm the improvement of the proposed algorithm.